\documentclass[a4paper,11pt]{article}
\usepackage{jcappub} % for details on the use of the package, please see the JINST-author-manual
\usepackage{lineno}
\usepackage{graphicx}
\UseRawInputEncoding

\title{\boldmath Restoring cosmological concordance with axion-like early dark energy and dark matter characterized by a constant equation of state?}

\author[a,*]{Yan-Hong Yao}

\author[b]{and Xin-He Meng}

\affiliation[a]{School of Physics and Astronomy, Sun Yat-sen University \\
2 Daxue Road, Tangjia, Zhuhai, People's Republic of China}

\affiliation[b]{School of Physics, Nankai University\\
94 Weijin Road, Nankai, Tianjin, People's Republic of China}

\emailAdd{yaoyh29@mail.sysu.edu.cn}

\abstract{The Hubble tension persists as a challenge in cosmology. Even early dark energy (EDE) models, initially considered the most promising for alleviating the Hubble tension, fall short of addressing the issue without exacerbating other tensions, such as the $S_8$ tension. Considering that a negative dark matter (DM) equation of state (EoS) parameter is conducive to reduce the value of $\sigma_8$ parameter, we extend the axion-like EDE model in this paper by replacing the cold dark matter (CDM) with DM characterized by a constant EoS $w_{\rm dm}$ (referred as WDM hereafter). We then impose constraints on this axion-like EDE extension model, along with three other models: the axion-like EDE model, $\Lambda$WDM, and $\Lambda$CDM. These constraints are derived from a comprehensive analysis incorporating data from the Planck 2018 cosmic microwave background (CMB), baryon acoustic oscillations (BAO), the Pantheon compilation, as well as a prior on $H_0$ (i.e., $H_0=73.04\pm1.04$, based on the latest local measurement by Riess et al.) and a Gaussianized prior on $S_8$ (i.e., $S_8=0.766\pm0.017$, determined through the joint analysis of KID1000+BOSS+2dLenS). We find that although the new model maintains the ability to alleviate the Hubble tension to $\sim$ 1.4$\sigma$, it still exacerbate the $S_8$ tension to a level similar to that of the axion-like EDE model.}

\begin{document}
\maketitle
\flushbottom

\section{Introduction}
\label{sec:intro}

Although scrutinized by a plethora of observational data across various scales in the past, the Lambda Cold Dark Matter ($\Lambda$CDM) model is currently facing skepticism regarding its internal inconsistencies. The Hubble tension, i.e. disagreement between direct~\cite{riess2022comprehensive} and indirect measurements~\cite{aghanim2020planck1} of the Hubble constant $H_0$, is a major factor contributing to this outcome. Although, up to this point, we can't exclude the possibility of systematics as the origin of the Hubble tension, many cosmologists are beginning to address the tension by introducing new physics. Generally speaking, solutions for resolving the Hubble tension in a theoretical manner can be categorized into two types: early-time solutions and late-time solutions, based on the introduction of new physics before and after recombination.  Early-time solutions, such as early dark energy (EDE) models~\cite{poulin2019early,agrawal2019rock,lin2019acoustic,ye2020hubble,smith2020oscillating,akarsu2020graduated,braglia2020unified,vagnozzi2021consistency,niedermann2021new,freese2021chain},
which introduce an exotic dark sector that acts as a cosmological constant before a critical redshift $z_c$ around 3000 but whose density then dilutes faster than radiation, and dark radiation models~\cite{battye2014evidence,Zhang2014Neutrinos,zhang2015sterile,feng2018searching,zhao2018measuring,choudhury2019constraining}, which include extra relativistic degrees of freedom that doesn't interact with photons and baryons, are considered to play a important role in resolving the Hubble tension. Nevertheless, as pointed out by the author of Ref.~\cite{vagnozzi2023seven}, early-time new physics alone will always fall short of fully solving the Hubble tension. On the other hand,  late-time solutions, such as late dark energy (DE) models~\cite{huang2016dark,vagnozzi2018constraints,martinelli2019cmb,visinelli2019revisiting,vagnozzi2020new,alestas2020h,alestas2021late}
and interacting DE models~\cite{kumar2016probing,kumar2017echo,di2017can,yang2018tale,yang2018interacting,kumar2019dark,yang2019dark,yao2020new1,di2020nonminimal,di2020interacting,lucca2020tensions,yang2020dynamical,nunes2022new},
are also considered to be incapable of fully resolving the Hubble tension, since consistency with Baryon Acoustic Oscillation (BAO) and uncalibrated Type Ia supernovae (SN Ia) data requires new physics to be introduced before recombination, in order to reduce the sound horizon by $\sim$ 7\%~\cite{bernal2016trouble,addison2018elucidating,lemos2019model,aylor2019sounds,knox2020hubble}.

Among all early-time solutions, EDE models may be the most promising category. However, similar to other early-time solutions, EDE models is incapable of fully resolving the Hubble tension, one of the primary reasons for this is that considering a non-zero EDE fractions $f_{\rm EDE}(z_c)$ near the matter-radiation equality would increase the dark matter (DM) density $\omega_{\rm cdm}$ compared to that of $\Lambda$CDM. A higher value of $\omega_{\rm cdm}$ would lead to a higher value of the $S_8=\sigma_8\sqrt{\Omega_{\rm m}/0.3}$ parameter (where $\Omega_{\rm m}$ is the matter density parameter and $\sigma_8$ is the matter fluctuation amplitude on scales of $ 8h^{-1}{\rm Mpc}$) if the value of $\sigma_8$ undergoes little variation, leading to a more significant $S_8$ tension between Cosmic Microwave Background (CMB) data and weak lensing (WL) as well as Large-Scale Structure (LSS) data~\cite{macaulay2013lower,joudaki2016cfhtlens,bull2016beyond,joudaki2017kids,nesseris2017tension,kazantzidis2018evolution,asgari2020kids+,hildebrandt2020kids+,skara2020tension,abbott2020dark,joudaki2020kids+,heymans2021kids,asgari2021kids,loureiro2021kids,abbott2022dark,amon2022dark,secco2022dark,philcox2022boss}.
However, it is still possible that an extension of an EDE model would mitigate the need for an increase in the $\omega_{\rm cdm}$, or alternatively compensate the associated increase in the amplitude of fluctuations. In fact, various extensions of EDE models have been proposed by researchers guided by this idea in order to simultaneously alleviate the $H_0$ tension and the $S_8$ tension. See e.g. Refs.~\cite{murgia2021early,allali2021dark,fondi2022no,karwal2022chameleon,mcdonough2022early,wang2022fraction,clark2023h,reeves2023restoring}. These models employ various methods to supplement or modify the cold dark matter (CDM) paradigm, including the introduction of the total neutrino mass $M_{\nu}$ as a free parameter~\cite{murgia2021early,fondi2022no,reeves2023restoring}, the inclusion of ultra-light axions that comprises five percent of DM~\cite{allali2021dark}, the proposal to replace CDM with decaying DM~\cite{clark2023h}, the consideration of DM coupled with EDE~\cite{karwal2022chameleon,mcdonough2022early,wang2022fraction}. Since the nature of DM remains mysterious, it is appropriate to explore phenomenological models and observe how they align with data. The simplest phenomenological modification to the CDM paradigm, i.e., considering a constant DM equation of state (EoS) as a free parameter, has not yet been explored in the existing literature related to the extension of EDE. Nevertheless, it is well known that a negative value of DM EoS has the capability to reduce the value of $\sigma_8$ parameter compared to a CDM setting. Therefore, in this paper, we will extend the axion-like EDE model by replacing CDM with DM characterized by a constant EoS (referred as WDM hereafter), and examine whether this extension can simultaneously alleviate the $H_0$ tension and the $S_8$ tension.

The rest of this paper is organized as follows. In section~\ref{sec:1}, we review the dynamics of axion-like EDE and WDM, and discuss their cosmological implications, specifically on the CMB and matter power spectra. In section~\ref{sec:2}, we describe the observational datasets and the statistical methodology. In section~\ref{sec:3}, we present the results of a Markov Chain Monte Carlo analysis applied on a combination of CMB, BAO, SN Ia data. In the last section, we make a brief conclusion for this paper.

\section{Model overview}\label{sec:1}

Our model consists of two modifications to $\Lambda$CDM. The first one is the inclusion of the axion-like EDE, and the second one is the replacement of CDM with WDM. Therefore we refer to this model as EDE+WDM hereafter. In this section, we will give a brief summary of the dynamics of EDE+WDM.

In EDE+WDM, EDE is represented by a canonical scalar field~\cite{smith2020oscillating}, while WDM is represented by an ideal fluid. Their energy density and pressure affect the dynamics of other species through Einstein's equation. At the homogeneous and isotropic level, the expansion rate of the universe can be written as:
\begin{equation}\label{}
  H=H_0\sqrt{\Omega_{\rm dm}(a)+\Omega_b(a)+\Omega_r(a)+\Omega_{\Lambda}+\Omega_{\phi}(a)},
\end{equation}
where $a$ is the scale factor, $\Omega_X\equiv\rho_{X}/\rho_{\rm crit}$, and $\rho_{\rm crit}=3H_0^2M_P^2$, with $M_P\equiv(8\pi G)^{-\frac{1}{2}}$ being the reduced Planck mass. The energy density and pressure of the scalar field at the background level are given by:
\begin{eqnarray}
% \nonumber to remove numbering (before each equation)
  \rho_{\phi} &=& \frac{1}{2}\dot{\phi}^2+V(\phi), \\
  p_{\phi} &=&  \frac{1}{2}\dot{\phi}^2-V(\phi),
\end{eqnarray}
where the dot indicates a derivative with respect to cosmic time. The potential of the axion-like EDE scalar field is chosen in the following form:
\begin{equation}\label{}
  V(\phi)=m^2f^2[1-\cos(\phi/f)]^3,
\end{equation}
where $m$ and $f$ are the EDE mass and decay constant. The evolution of the scalar field $\phi$ is described by the Klein-Gorden (KG) equation:
\begin{equation}\label{}
  \ddot{\phi}+3H\dot{\phi}+V_{,\phi}=0.
\end{equation}
As described in the literature~\cite{smith2020oscillating}, initially, the scalar field is frozen due to Hubble friction at a position displaced from the minimum of its potential, and its energy density is sub-dominant. It is only after the Hubble rate drops below the effective mass of the scalar field that the field becomes dynamical, starts rolling down, and oscillates around the minimum of its potential. This leads to a faster dilution of its energy density compared to radiation. Given the evolutionary behavior of EDE, the fundamental particle physics parameters $m$ and $f$ can be related to the phenomenological parameters $\log_{10}z_c$ and $f_{\text{EDE}}(z_c)$. Here $f_{\text{EDE}}(z_c)$ is the maximum fraction of the total energy density in the scalar field, and $z_c$ is the redshift at which this fraction reaches its maximum. Therefore, the EDE component is governed by three parameters: $\log_{10}z_c$, $f_{\text{EDE}}(z_c)$, and the initial misalignment angle $\theta_i=\phi_i/f$, with $\phi_i$ being the initial field value. As showed in~\cite{smith2020oscillating}, $m$ largely controls the value of $z_c$, while $f$ controls the value of $f_{\text{EDE}}(z_c)$. And the approximate equations related $m$ to $z_c$ and $f$ to $f_{\text{EDE}}(z_c)$ are:
\begin{equation}\label{}
  m^2|(1-\cos\theta_i)^{2}(2+3\cos\theta_i)| \simeq 3H(z_c)^2
\end{equation}
\begin{equation}\label{}
  f_{\text{EDE}}(z_c) \simeq \frac{m^2f^2}{\rho_{tot}(z_c)}(1-\cos\theta_i)^{3}
\end{equation}
It can be inferred from the above equations that for a fixed $\theta_i$ a value of $m$ determines $z_c$ and a value of $f$ determines $f_{\text{EDE}}(z_c)$. According to~\cite{karwal2016}, $\theta_i$, once other EDE parameters is fixed, is a parameter whose value controls the oscillation frequency of the background field, it is tightly constrained by the small-scale polarization measurements, in fact, the full Planck 2015 dataset excluding the region $\theta_i<1.8$ at 95\% confidence level.

The evolution of the energy density of WDM, however, is described by a simpler equation, namely the continuity equation of WDM:
\begin{equation}\label{}
  \dot{\rho}_{\rm dm}+3H(1+w_{\rm dm})\rho_{\rm dm}=0.
\end{equation}
Here, $w_{\rm dm}$ is the EoS of WDM, and it is a constant. After solving this equation, we obtain $\rho_{\rm dm}=\rho_{\rm dm0}a^{-3(1+w_{\rm dm})}$ (without losing any generality we set the current value of the scale factor $a_0$ to be unity). According to this formula, WDM dilutes either faster or slower than CDM, depending on the sign of $w_{\rm dm}$.

The EDE+WDM model is determined by ten parameters: six basic parameters shared with $\Lambda$CDM, plus four new parameters, namely $\log_{10}z_c$, $f_{\text{EDE}}(z_c)$, $\theta_i$, and $w_{\rm dm}$.

At the linear perturbation level, the perturbation of scalar field $\phi$ (in the Fourier space) is governed by the linearized KG equation:
\begin{equation}\label{}
  \delta\phi_k^{\prime\prime}+2\mathcal{H}\delta\phi_k^{\prime}+(k^2+a^2V_{,\phi\phi})\delta\phi_{k}=-h^{\prime}\phi^{\prime}/2
\end{equation}
where the prime denotes derivative with respect to conformal time, $\mathcal{H}$ is the conformal Hubble parameter, $h$ is the metric potential in synchronous gauge, and $k$ is magnitude of the wavenumber $\vec{k}$.

On the other hand, the linear perturbation equations of WDM is consist of the continuity and Euler equations of this component,i.e.:
\begin{equation}
 \delta^{\prime}_{\rm dm}= -(1+w_{\rm dm}) \left(\theta_{\rm dm}+\frac{1}{2}h^{\prime} \right)
 -3 \mathcal{H} \delta_{\rm dm} (c^2_{\rm s,dm} - w_{\rm dm}) - 9 (1+w_{\rm dm})(c^2_{\rm s,dm} - c^2_{\rm a,dm})\mathcal{H}^2 \frac{\theta_{\rm dm}}{k^2},
\end{equation}
\begin{equation}
\theta^{\prime}_{\rm dm}=-(1-3 c^2_{\rm s,dm}) \mathcal{H} \theta_{\rm dm}  + \frac{c^2_{\rm s,dm}}{1+w_{\rm dm}}k^2 \delta_{\rm dm}.
\end{equation}
where $\delta_{\rm dm}$ and $\theta_{\rm dm}$ are the relative density and velocity divergence perturbations of DM, $c_{\rm s,dm}$ and $c_{\rm a,dm}=\dot{p}_{\rm dm}/\dot{\rho}_{\rm dm}$ denote the sound speed and adiabatic sound speed of DM. The sound speed of DM describe its micro-scale properties and needs to be provided independently, in this article, we consider $c^2_{\rm s,dm}=0$ in order to understand the extent to which modifying the DM EoS alone can alleviate the $S_8$ tension. Having presented the equations above, the background and perturbation dynamics of the EDE+WDM model is clearly understood.

\begin{figure*}
%\begin{tabular}{cc}
\begin{minipage}{0.49\linewidth}
  \centerline{\includegraphics[width=1\textwidth]{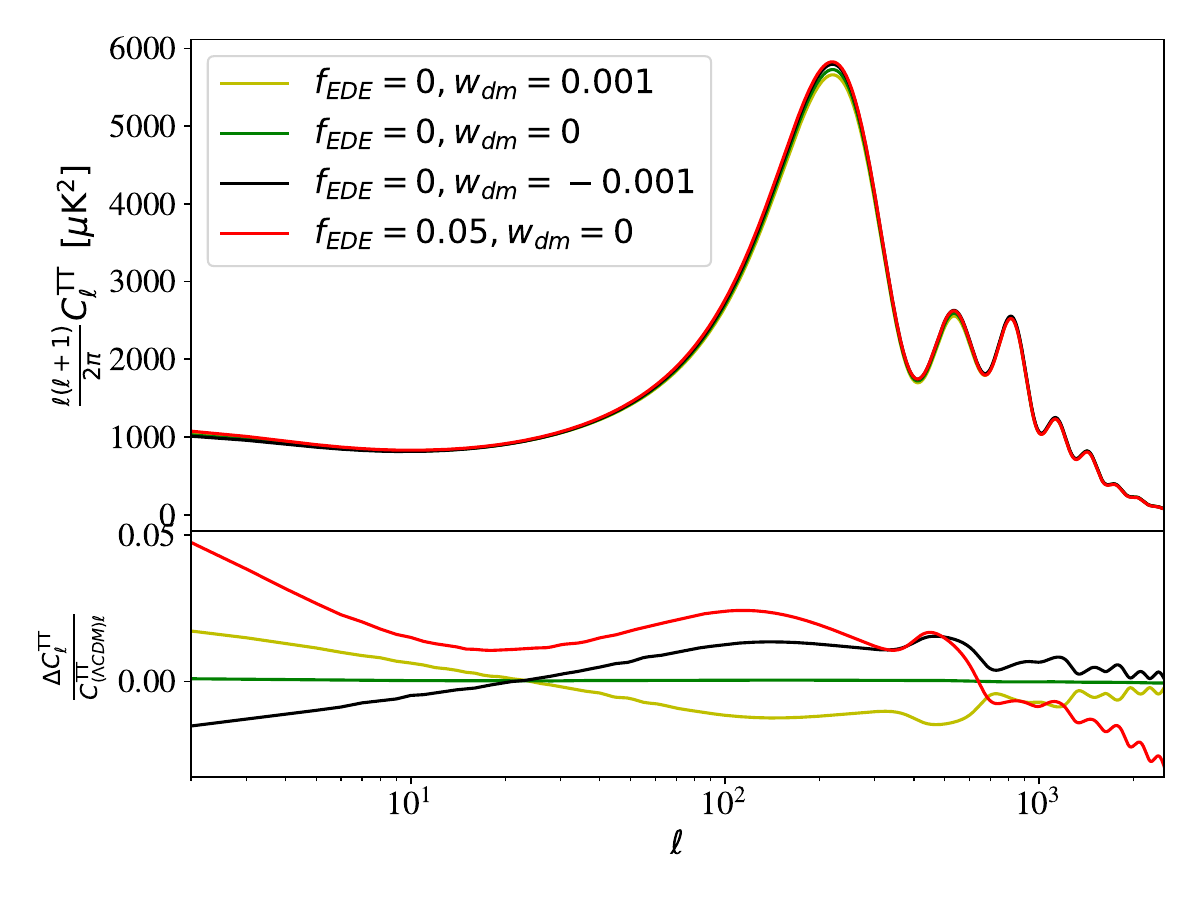}}
\end{minipage}
\begin{minipage}{0.49\linewidth}
  \centerline{\includegraphics[width=1\textwidth]{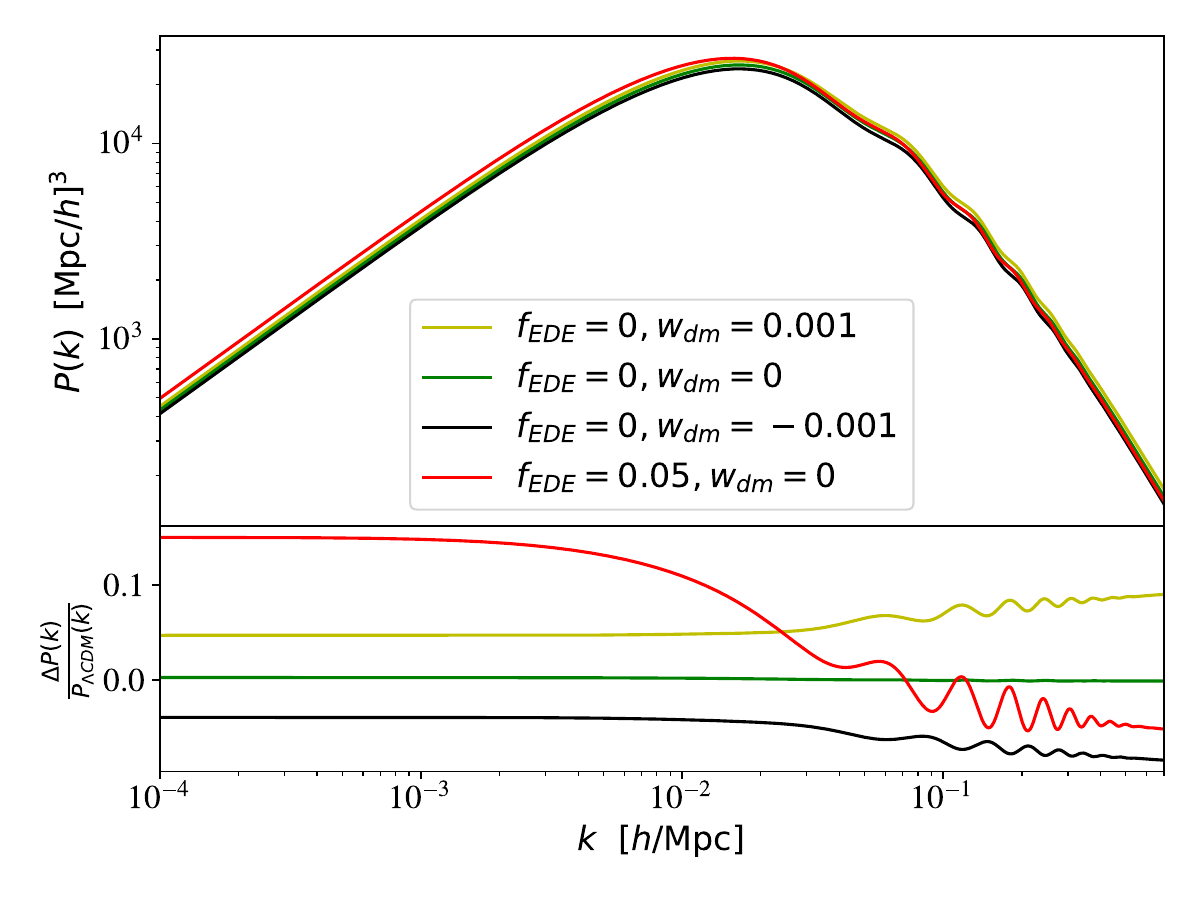}}
\end{minipage}
\caption{The CMB TT and matter power spectra are computed for different values of the parameters $f_{\text{EDE}}(z_c)$ and $w_{dm}$, while keeping six basic parameters fixed at the values of their counterparts in the baseline $\Lambda$CDM model, as extracted from the Planck 2018 + lowE + lensing data. Additionally, $\log_{10}z_c$ and $\theta_i$ are set to 3.5 and 2.5.}
\label{fig:0}
\end{figure*}

At the end of this section, we present some implications regarding the impacts of EDE+WDM on the CMB TT and matter power spectra. For a simplified demonstration of these effects, we display in Fig.\ref{fig:0} the CMB TT and matter power spectra of EDE+WDM, along with their residuals compared to a baseline $\Lambda$CDM model. We define our benchmark $\Lambda$CDM model with the following cosmological parameters: the peak scale parameter 100$\theta_s$=1.041783, the baryon density today $\omega_b$=0.02238280, the DM density today $\omega_{\rm dm}$=0.1201705, the optical depth $\tau_{\rm reio}$=0.0543082, the amplitude of the primordial scalar $A_s=2.100549\times10^{-9}$, the spectral index of the primordial scalar $n_s$=0.9660499. These values are extracted from the Planck 2018 + lowE + lensing dataset~\cite{aghanim2020planck1}. The six basic parameters of EDE+WDM are fixed to the values of their counterparts in the baseline $\Lambda$CDM model. Other parameters, namely $\log_{10}z_c$, $f_{\text{EDE}}(z_c)$, $\theta_i$, and $w_{\text{dm}}$, are set to the following values for different choices: 3.5, 0, 2.5, 0.001; 3.5, 0, 2.5, 0; 3.5, 0, 2.5, $-0.001$; 3.5, 0.05, 2.5, 0. Of course, we
point out that the values of six basic parameters of EDE+WDM extracted from Planck 2018 + lowE + lensing dataset are different from the six corresponding values provided above, in fact, EDE+WDM have a larger value of $\omega_{\rm cdm}$ compared to $\Lambda$CDM due to the $w_{\text{dm}}$-$\omega_{\rm dm}$ and EDE-$\omega_{\rm dm}$ degeneracy, and this will lead to different CMB TT and matter power spectra. However, in order to consider the individual impact of changing the values of $w_{\text{dm}}$ and $f_{\text{EDE}}(z_c)$ on the CMB TT and matter power spectra, we set the values of the six basic parameters for both EDE+WDM and $\Lambda$CDM to be consistent.

Focusing on the CMB TT power spectrum and fixing the $f_{\text{EDE}}(z_c)$ parameter, we observe that the amplitudes of the acoustic peaks in the CMB predicted by EDE+WDM with a negative $w_{\rm dm}$ are increased. This is attributed to the fact that negative values of the parameter $w_{\rm dm}$ will postpone the moment of matter-radiation equality when other model parameters are fixed. On larger scales where $l<10$, the curves are depressed when the value of the parameter $w_{\rm dm}$ is negative, due to the integrated Sachs-Wolfe effect. In addition, when the values of the parameter $w_{\rm dm}$ turn positive, while keeping the values of other relevant cosmological parameters unchanged, the resulting effects are opposite to those observed with negative values of $w_{\rm dm}$. On the other hand, when fixing the $w_{\rm dm}$ parameter, an increase in the amplitudes of the acoustic peaks in the CMB is observed in EDE+WDM when $f_{\text{EDE}}(z_c)$ is non-zero. Additionally, the curves show an increase on larger scales under this circumstance. These effects differ from those arising from a positive or negative $w_{\rm dm}$.

Concentrating on the matter power spectrum with a fixed $f_{\text{EDE}}(z_c)$ parameter, it becomes evident that a negative value of $w_{\rm dm}$ results in a decrease, while positive values of $w_{\rm dm}$ lead to an increase in the matter power spectrum, this is because the former situation results in a delay of matter and radiation equality, while the latter situation leads to an advance of matter and radiation equality. On the other hand, when fixing the $w_{\rm dm}$ parameter, a non-zero value of $f_{\text{EDE}}(z_c)$ leads to a different outcome, i.e. an increase in the large scale and a decrease in the small scale of the matter power spectrum. From these effects, one can infer that a non-zero $f_{\text{EDE}}(z_c)$ itself is not the direct reason for worsening the $S_8$ tension, considering it leads to a decrease in the small scale of the power spectrum, the problem arise from the EDE-$\omega_{\rm dm}$ degeneracy, it counteracts the effect of a non-zero $f_{\text{EDE}}(z_c)$, and we hope that a negative $w_{\rm dm}$, in addition to a non-zero $f_{\text{EDE}}(z_c)$, will solve this problem\footnote{It's worth mentioning that, if both the values of $w_{\rm dm}$ and $f_{\text{EDE}}(z_c)$ are non-zero, then it's not difficult to infer that, as long as both values are small enough, the observed results in the power spectrum will be a combination of the outcomes corresponding to only one of the two values being zero.}.

\section{Datasets and methodology}\label{sec:2}
In order to extract the mean values and confidence intervals of the model parameters, we utilize the recent observational datasets described below.

\textbf{Cosmic Microwave Background (CMB)}: we make use of the Planck 2018~\cite{aghanim2020planck1,aghanim2020planck2,aghanim2020planck3} CMB temperature, polarization, and lensing measurements that includes plik TTTEEE, lowl, lowE, and lensing likelihood.

\textbf{Baryon acoustic oscillations(BAO)}: we also utilize several BAO distance measurements including 6dFGS~\cite{Beutler2011The}, SDSS-MGS~\cite{ross2015clustering}, and BOSS DR12~\cite{alam2017clustering}.

\textbf{Pantheon}: the Pantheon catalogue of Supernovae Type Ia, comprising 1048 data points in the redshift region $z\in[0.01, 2.3]$ are also considered.

\textbf{Hubble constant}: the latest local measurement of Hubble constant obtained by Riess et al.~\cite{riess2022comprehensive}, i.e. $H_0=73.04\pm1.04$ are also included, we denoted it as R22 hereafter.

\textbf{$S_8$ parameter}: in addition to the above datasets, we include a Gaussianized prior on $S_8$, i.e.~$S_8=0.766\pm0.017$, chosen according to the joint analysis of KID1000+BOSS+2dLenS. (The use of a prior as an approximation for the full WL likelihoods has been demonstrated to be justified in the context of EDE models~\cite{hill2020early}. Nevertheless, in the case of the EDE+WDM model, a comprehensive assessment of the likelihoods necessitates a dedicated treatment of nonlinearities. Due to the absence of such tools, we restrict our analysis to the linear power spectrum, and make the assumption that the incorporation of this $S_8$ prior correctly captures the constraints from the KID1000+BOSS+2dLenS likelihoods on EDE+WDM.)

To constrain the EDE+WDM model, we run a Markov Chain Monte Carlo (MCMC) using the public code MontePython-v3~\cite{audren2013conservative,brinckmann2019montepython}, interfaced with a modified version of the CLASS\_EDE code~\cite{hill2020early} which is an extension to the CLASS code~\cite{lesgourgues2011cosmic,blas2011cosmic}. We perform the analysis with a Metropolis-Hasting algorithm and consider chains to be converged using the Gelman-Rubin~\cite{gelman1992inference} criterion $R-1<0.03$, assuming flat priors on the following parameter space:
\begin{equation}\label{}
  \mathcal{P}=\{\omega_b, \omega_{\rm dm}, \theta_s, A_s, n_s, \tau_{\rm reio}, \log_{10}z_c, f_{\rm EDE}(z_c), \theta_i, w_{\rm dm}\},
\end{equation}
we also adopt the Planck collaboration convention and model free-streaming neutrinos as two massless species and one massive with $M_v=0.06$~eV.
\section{Results and discussion}\label{sec:3}

In this section we constrain EDE+WDM, EDE\footnote{'EDE' here represents the axion-like early dark energy model instead of early dark energy, the specific meaning of 'EDE' will no longer be explicitly stated in the following content. Readers are encouraged to infer its significance based on the context.}, $\Lambda$WDM, $\Lambda$CDM using CMB, BAO, Pantheon datasets, as well as R22 and the Gaussianized prior on $S_8$ that we shown in the previous section in order to perform a statistical comparison between these models with the aim to focus on the tension on both $H_0$ and $S_8$.

\begin{figure*}
	\centering
	\includegraphics[scale=0.4]{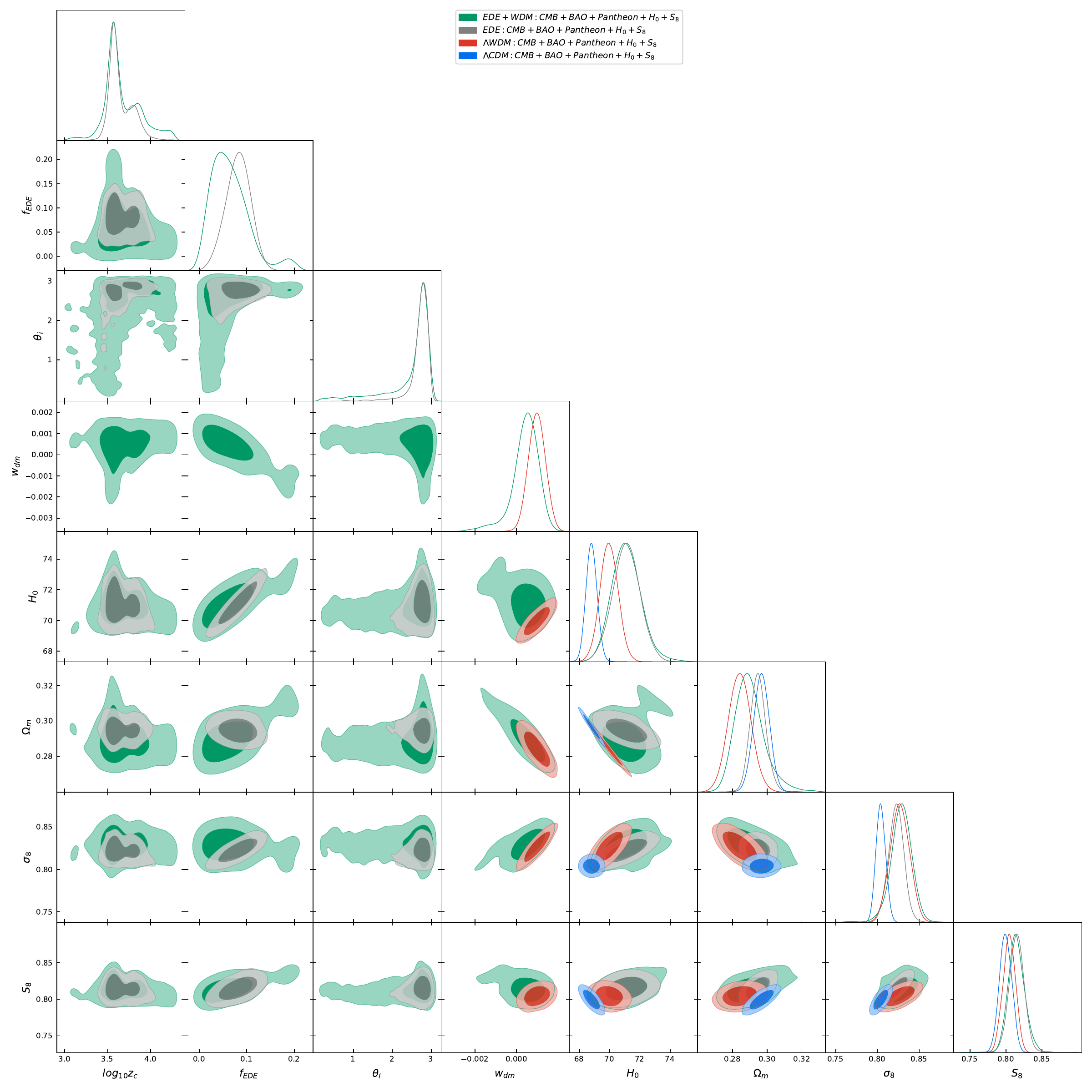}
	\caption{one dimensional posterior distributions and two dimensional joint contours at 68\% and 95\% confidence levels for the most relevant parameters of EDE+WDM, EDE, $\Lambda$WDM, and $\Lambda$CDM are presented using CMB+BAO+Pantheon+$H_0$+$S_8$ dataset. }
\label{fig:1}
\end{figure*}

\begin{figure*}
	\centering
	\includegraphics[scale=0.8]{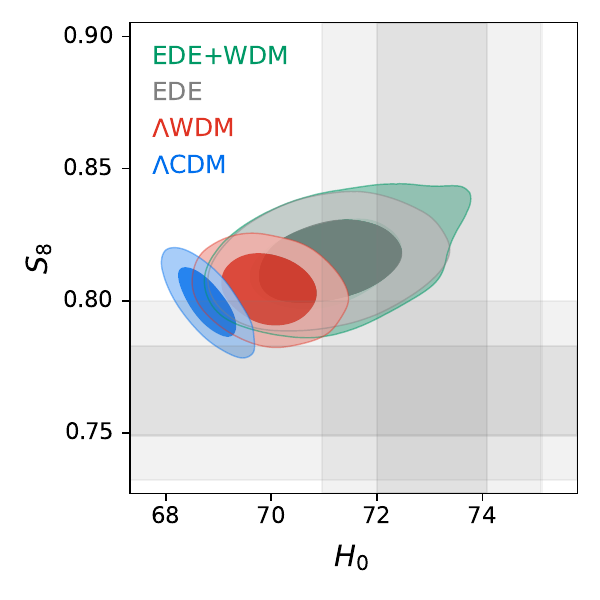}
	\caption{the $S_8$-$H_0$ contours at 68\% and 95\% confidence levels for EDE+WDM, EDE, $\Lambda$WDM, and $\Lambda$CDM regarding CMB+BAO+Pantheon+$H_0$+$S_8$ dataset, as well as the boundaries corresponding to one standard deviation for R22 and the Gaussianized prior.}
\label{fig:2}
\end{figure*}

In Table~\ref{tab:1}, we present the observational constraints on EDE+WDM, EDE, $\Lambda$WDM, and $\Lambda$CDM based on CMB+BAO+Pantheon+$H_0$+$S_8$ dataset.
Fig.~\ref{fig:1} shows the one dimensional posterior distributions and two dimensional joint contours at 68\% and 95\% confidence levels for the most relevant parameters of EDE+WDM, EDE, $\Lambda$WDM, and $\Lambda$CDM.  From Table~\ref{tab:1}, one can see that the values of parameters $w_{\rm dm}$ of both EDE+WDM and $\Lambda$WDM are very close to 0. This doesn't come as a surprise, since otherwise the large-scale structure can not be correctly formed. In addition, we find that the fitting results for $H_0$ parameter in EDE+WDM, i.e. $71.11^{+0.88}_{-1.0}$ at $68\%$ confidence level, is similar to that of EDE, i.e. $71.10\pm 0.90$ at $68\%$ confidence level. It still maintains the ability to alleviate the Hubble tension to $\sim$ 1.4 $\sigma$, which is a notable improvement compared to that of $\Lambda$WDM, i.e. $69.95\pm 0.60$ at $68\%$ confidence level (with Hubble tension $\sim$ 2.6 $\sigma$) and that of $\Lambda$CDM, i.e. $68.79\pm 0.37$ at $68\%$ confidence level (with Hubble tension $\sim$ 3.9 $\sigma$). However, unfortunately, we find that the fitting result for $S_8$ parameter in EDE+WDM, i.e. $0.814^{+0.011}_{-0.012}$ at $68\%$ confidence level is also similar to that of EDE, i.e. $0.815\pm 0.011$ at $68\%$ confidence level, showing a $\sim$ 2.3 $\sigma$ tension on the $S_8$ parameter, which is still worse than that of $\Lambda$WDM, i.e. $S_8=0.8043^{+0.0093}_{-0.0084}$ $68\%$ confidence level (with $S_8$ tension $\sim$ 2 $\sigma$) and that of $\Lambda$CDM, i.e. $S_8=0.7994\pm 0.0086$ $68\%$ confidence level (with $S_8$ tension $\sim$ 1.8 $\sigma$). We attribute the failure of the EDE+WDM model to alleviate the $S_8$ tension to the lack of a positive (negative) correlation between the $S_8$ parameter and the $w_{\rm dm}$ parameter, as well as a sufficiently large negative (positive) value of $w_{\rm dm}$ in this model. Although, as expected, there is a positive correlation between the $w_{\rm dm}$ parameter and the $\sigma_8$ parameter in this model, the negative correlation between the $w_{\rm dm}$ parameter and $\Omega_m$ leads to the lack of correlation between the $S_8$ parameter and the $w_{\rm dm}$ parameter. This is somewhat different from the situation in $\Lambda$WDM. Although in $\Lambda$WDM, the $w_{\rm dm}$ parameter is also positively correlated with the $\sigma_8$ parameter while being negatively correlated with the $\Omega_m$ parameter, it still results in a slight positive correlation between the $w_{\rm dm}$ parameter and the $S_8$ parameter. We note that even though there is a slight positive correlation between the $w_{\rm dm}$ parameter and the $S_8$ parameter in the $\Lambda$WDM case, such a model still exacerbates the $S_8$ tension compared to $\Lambda$CDM due to the positive value of $w_{\rm dm}$. To visually illustrate the tension among the fitting results of EDE+WDM, EDE, $\Lambda$WDM, $\Lambda$CDM, and two priors included in the datasets (R22 and the Gaussianized prior on $S_8$), Fig.~\ref{fig:2} reproduces the $S_8$-$H_0$ contours, incorporating the boundaries corresponding to one standard deviation for these two priors. It's worth mentioning that, the above result is still obtained when using both $H_0$ prior and $S_8$ prior simultaneously. If we were to use only one of these priors, or none at all, the EDE+WDM model would face even greater Hubble and $S_8$ tensions.

We also consider the Akaike information criterion (AIC) for model comparison among EDE+WDM, EDE, $\Lambda$WDM, and $\Lambda$CDM. The AIC is defined as $\chi_{\text{min}}^2 + 2k$, where $k$ denotes the number of cosmological parameters. In practice, we are primarily interested in the relative values of AIC between two different models, denoted as $\Delta$AIC = $\Delta\chi_{\text{min}}^2 + 2\Delta k$. A model with a smaller AIC value is considered to be more favored. In this work, the $\Lambda$CDM model serves as the reference model. From Table \ref{tab:1}, we find that the $\Delta$AIC values for EDE+WDM, EDE, and $\Lambda$WDM are $-5.2$, $-7.18$, $-4.14$, respectively. Therefore, for the CMB+BAO+Pantheon+$H_0$+$S_8$ dataset, EDE is the most favored model among the four models analyzed in this work.

Generally speaking, the observational data force $w_{\rm dm}$ of EDE+WDM to be very close to 0, resulting in EDE+WDM not differing much from EDE. Additionally, a negative $w_{\rm dm}$ lacks a common physical explanation. Also, as shown above, EDE+WDM is not favored by the observational data compared to EDE, as its AIC is worse than that of EDE. Furthermore, EDE+WDM cannot alleviate the Hubble tension and the $S_8$ tension. In short, adding a new degree of freedom $w_{\rm dm}$ is not a good choice.

\begin{table*}[ht]
    \centering
    \scalebox{0.8}[0.8]{
    \begin{tabular}{|c|c|c|c|c|}
        \hline
        Model & ~$\Lambda$CDM~ & ~$\Lambda$WDM~ &~ EDE~ & ~EDE+WDM~ \\ \hline
        Dataset & \multicolumn{4}{c|}{~CMB + BAO + Pantheon + $S_8$ + $H_0$~}  \\ \hline \hline
        $100~\omega_{b}$ & $2.265\pm 0.013^{+0.025}_{-0.025} $ & $2.248\pm 0.014^{+0.027}_{-0.027}$ & $2.286\pm 0.021^{+0.041}_{-0.040}$ & $2.274\pm 0.024^{+0.048}_{-0.045} $ \\
        $\omega_{\rm cdm}$ & $0.11707\pm 0.00081^{+0.0016}_{-0.0016}$ & $0.11586\pm 0.00089^{+0.0018}_{-0.0017}$ & $0.1255\pm 0.0033^{+0.0063}_{-0.0062}$ & $ {0.1233^{+0.0027}_{-0.0060}}^{+0.020}_{-0.0088}$ \\
        $100~\theta_{s}$ & $1.04218\pm 0.00029^{+0.00055}_{-0.00058}$ & $1.04202\pm 0.00028^{+0.00055}_{-0.00056}$ & $1.04163\pm 0.00038^{+0.00071}_{-0.00073}$ & ${1.04163^{+0.00042}_{-0.00031}}^{+0.00082}_{-0.0012} $ \\
        $\ln(10^{10}A_{s})$ & $3.050\pm 0.015^{+0.030}_{-0.028}$ & $3.045\pm 0.014^{+0.028}_{-0.028}$ & $3.057\pm 0.015^{+0.029}_{-0.029}$ & $3.053\pm 0.015^{+0.032}_{-0.030} $\\
        $n_{s}$  & $0.9725\pm 0.0037^{+0.0071}_{-0.0072}$ & $0.9701\pm 0.0036^{+0.0071}_{-0.0072}$ & $0.9859\pm 0.0065^{+0.013}_{-0.013}$ & ${0.9818^{+0.0068}_{-0.010}}^{+0.022}_{-0.018} $ \\
        $\tau_\mathrm{reio}$ & $0.0598\pm 0.0075^{+0.016}_{-0.014}$ & ${0.0551^{+0.0066}_{-0.0075}}^{+0.015}_{-0.014}$ & $0.0571\pm 0.0075^{+0.015}_{-0.015}$ & $0.0556\pm 0.0073^{+0.015}_{-0.014} $ \\
        $\log_{10}z_c$ & $-$ & $-$ & $3.65\pm 0.14^{+0.29}_{-0.22}$ & $3.67\pm 0.22^{+0.54}_{-0.37}  $ \\
        $ f_{\rm EDE}(z_c)$ & $-$ & $-$ & ${0.082^{+0.028}_{-0.025}}^{+0.049}_{-0.054}$ & ${0.066^{+0.025}_{-0.047}}^{+0.13}_{-0.071} $ \\
        $ \theta_i$ & $-$ & $-$ & ${2.71^{+0.23}_{-0.069}}^{+0.34}_{-0.47}$ & ${2.48^{+0.53}_{+0.027}}^{+0.62}_{-1.6} $ \\
        $ w_{\rm dm}$ & $-$ & $0.00098\pm 0.00041^{+0.00079}_{-0.00079}$ & $-$ & $ {0.00042^{+0.00069}_{-0.00043}}^{+0.0013}_{-0.0014}$ \\
        $ H_0$ & $68.79\pm 0.37^{+0.73}_{-0.70} $ & $69.95\pm 0.60^{+1.2}_{-1.2} $ & $71.10\pm 0.90^{+1.8}_{-1.8}$ & ${71.11^{+0.88}_{-1.0}}^{+1.9}_{-1.9} $\\
        $\Omega_{m}$& $0.2966\pm 0.0046^{+0.0090}_{-0.0091}$ & $0.2841\pm 0.0065^{+0.013}_{-0.013}$ & ${0.2948^{+0.0042}_{-0.0048}}^{+0.0096}_{-0.0087}$ & ${0.2901^{+0.0064}_{-0.0097}}^{+0.019}_{-0.016}$\\
        $ \sigma_8 $& $0.8039\pm 0.0057^{+0.011}_{-0.011}$ & $0.827\pm 0.011^{+0.022}_{-0.023} $ & $0.8224\pm 0.0094^{+0.018}_{-0.018}$ & $0.828\pm 0.014^{+0.025}_{-0.026} $ \\
        $ S_8$ & $0.7994\pm 0.0086^{+0.017}_{-0.017}$ & ${0.8043^{+0.0093}_{-0.0084}}^{+0.017}_{-0.017} $ & $0.815\pm 0.011^{+0.021}_{-0.021} $ & ${0.814^{+0.011}_{-0.012}}^{+0.025}_{-0.022} $\\ \hline
        $\chi_{min}^2$ &3840.76& 3834.62&3827.58&3827.56\\
        $\Delta\chi_{min}^2$ &$0$& $-6.14$&$-13.18$&$-13.20$\\
        $k$ & 28 & 29 &31  &32 \\
        AIC &3896.76  &3892.62  &3889.58  &3891.56 \\
        $\Delta$AIC & $0$ & $-4.14 $ & $-7.18 $ & $-5.2 $  \\ \hline
    \end{tabular}}
    \caption{The mean values and 1, 2 $\sigma$ errors of the parameters for EDE+WDM, EDE, $\Lambda$WDM, and $\Lambda$CDM are provided for the CMB+BAO+Pantheon+$S_8$+$H_0$ dataset, along with the AIC values for these four models.}
    \label{tab:1}
\end{table*}


\section{concluding remarks} \label{sec:4}

The Hubble tension remains a complex issue in cosmology, the current consensus within the scientific community suggests that, if systematics are not the origin of the Hubble tension, modifications are necessary both in the early and late stages compared to $\Lambda$CDM~\cite{vagnozzi2023seven}.
Early-time modification are deemed necessary to simultaneously increase the Hubble constant and reduce the sound horizon. However, such modifications fail to address the Hubble tension without exacerbating other tensions, such as the $S_8$ tension. One of the most prominent early-time solutions is the axion-like EDE model, which considers a non-zero EDE fraction $f_{\rm EDE}(z_c)$ near the matter-radiation equality. This increases the DM density and exacerbates the $S_8$ tension. Given that the axion-like EDE model is a typical early-time solution and a negative DM EoS is conducive to reduce the value of $\sigma_8$ parameter, we extend the axion-like EDE model by replacing the CDM in this model with DM characterized by a constant EoS $w_{\rm dm}$. We then impose constraints on this axion-like EDE extension model, i.e. EDE+WDM, along with three other models: the axion-like EDE model, i.e. EDE, $\Lambda$WDM, and $\Lambda$CDM. These constraints are extracted from a comprehensive analysis incorporating data from the Planck 2018 CMB, BAO, the Pantheon compilation, as well as R22 and the Gaussianized prior on $S_8$ determined through the joint analysis of KID1000+BOSS+2dLenS. Our findings indicate that, while EDE+WDM maintains EDE's ability to alleviate the Hubble tension to $\sim$ 1.4$\sigma$, it still moderately exacerbate the $S_8$ tension. In fact, a non-zero DM EoS not only affects the value of $\sigma_8$ but also shifts the value of $\Omega_m$; the combination of these two effects leads to the value of $S_8$ remaining close to that of the CDM case. Finally, we employ AIC to make a model comparison between EDE+WDM, EDE, $\Lambda$WDM, and $\Lambda$CDM regarding the CMB+BAO+Pantheon+$H_0$+$S_8$ dataset. Our analysis reveals that the EDE model is the most supported model among these four models regarding this dataset.  Since EDE+WDM is not favored by the observational data compared to EDE, and EDE+WDM cannot alleviate both the Hubble tension and the $S_8$ tension, we conclude that adding a new degree of freedom $w_{\rm dm}$ is not a good choice.

\acknowledgments

This work is supported by the National key R\&D Program of China (Grant No. 2020YFC2201600), National Natural Science Foundation of China (NSFC) under Grant No. 12073088, and National SKA Program of China No. 2020SKA0110402.

% Bibliography

%% [A] Recommended: using JHEP.bst file
%% \bibliographystyle{JHEP}
%% \bibliography{biblio.bib}

%% or
%% [B] Manual formatting (see below)
%% (i) We suggest to always provide author, title and journal data or doi:
%% in short all the informations that clearly identify a document.
%% (ii) please avoid comments such as "For a review'', "For some examples",
%% "and references therein" or move them in the text. In general, please leave only references in the bibliography and move all
%% accessory text in footnotes.
%% (iii) Also, please have only one work for each \bibitem.

\providecommand{\href}[2]{#2}\begingroup\raggedright\endgroup

\end{document}